\documentclass[journal=jacsat,manuscript=article]{achemso}

\usepackage[version=3]{mhchem} 
\usepackage{booktabs}
\usepackage{hyperref}
\usepackage{caption, subcaption}

\usepackage{pythonhighlight}

\SectionNumbersOn
\author{Shiqiang Zhang}
\affiliation[Imperial]{Imperial College London, London, UK}
\author{Christian W. Feldmann}
\affiliation[BASF]{BASF SE, Ludwigshafen, Germany}
\author{Frederik Sandfort}
\affiliation[BASF]{BASF SE, Ludwigshafen, Germany}
\author{Miriam Mathea}
\affiliation[BASF]{BASF SE, Ludwigshafen, Germany}
\author{Juan S. Campos}
\affiliation[Imperial]{Imperial College London, London, UK}
\author{Ruth Misener}
\affiliation[Imperial]{Imperial College London, London, UK}

\email{r.misener@imperial.ac.uk}

\title{Limeade: Let integer molecular encoding aid}

\abbreviations{IR,NMR,UV}

\begin{document}







\begin{abstract}
    Mixed-integer programming (MIP) is a well-established framework for computer-aided molecular design (CAMD). By precisely encoding the molecular space and score functions, e.g., a graph neural network, the molecular design problem is represented and solved as an optimization problem, the solution of which corresponds to a molecule with optimal score. However, both the extremely large search space and complicated scoring process limit the use of MIP-based CAMD to specific and tiny problems. Moreover, optimal molecule may not be meaningful in practice if scores are imperfect. Instead of pursuing optimality, this paper exploits the ability of MIP in molecular generation and proposes Limeade as an end-to-end tool from real-world needs to feasible molecules. Beyond the basic constraints for structural feasibility, Limeade supports inclusion and exclusion of SMARTS patterns, automating the process of interpreting and formulating chemical requirements to mathematical constraints. 
\end{abstract}

\section{Introduction}\label{sec:introduction}
Computer-aided molecular design (CAMD) \cite{Gani2004,Ng2014,Austin2016,Gani2022} aims to explore the massive molecular space and discover new products with desired properties. As a long-standing challenge, countless methods have been developed in this field from various perspectives. Inherit from \citeauthor{Alshehri2020}\cite{Alshehri2020}, we divide these methods into knowledge-driven methods and data-driven methods. Relying on semi-empirical quantitative structure-property relationships (QSPRs), knowledge-driven CAMD characterizes and ranks the design space, then finds potential candidates using optimization algorithms \cite{Odele1993,Churi1996,Camarda1999,Sinha1999,Sahinidis2003,Zhang2015,Liu2019,Alshehri2020,Hatamleh2022}. Instead of using knowledge from experts or statistics, data-driven CAMD automatically learns structure-property relationships from data. With the rapid development and outstanding performance of machine learning (ML) models, ML surrogates, especially graph neural networks (GNNs) \cite{Wu2020,Zhou2020,Chami2022,Rittig2023,Gao2024}, are widely used to learn the structure-property relationships and predict molecular properties \cite{Xu2017,Wang2019,Yang2019,Schweidtmann2020,Withnall2020,Alshehri2021,Sharma2021,Chen2023,Mann2023,Tiew2023,Dobbelaere2024}. Beyond prediction, the generation and optimization of molecules also benefit both from coupling ML with optimization techniques \cite{Gaudelet2021,Xiong2021,Rittig2022,Fromer2023} and from advanced deep generative modeling \cite{Lima2016,Elton2019,Xia2019,Alshehri2020,Faez2021,Mouchlis2021,Pirnay2024,Tang2024}. 

\begin{figure}[t]
    \centering
    \includegraphics[width=0.8\linewidth]{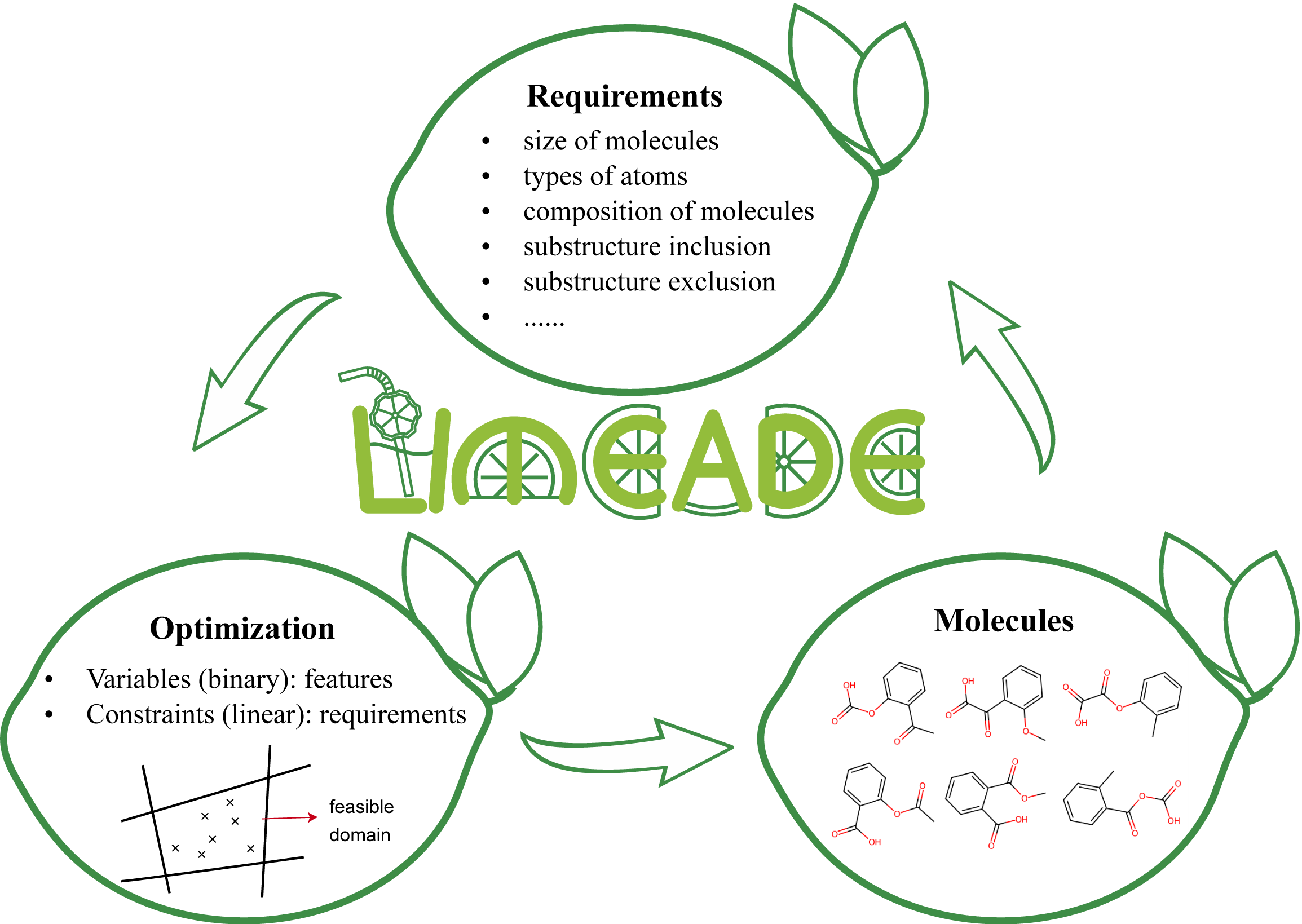}
    \caption{Limeade takes practical requirements as inputs, encodes molecular space and requirements into linear constraints over binary variables, using MIP solver to generate feasible solutions, and finally decodes solutions into molecules as outputs. After observing generated molecules, users might add more requirements and use Limeade to generate again.}
    \label{fig:Limeade}
\end{figure}

A general CAMD method usually consists of the following blocks: (i) molecular representation, (ii) property prediction, (iii) molecule generation, (iv) optimization. There are multiple molecular representations, e.g., fingerprints, SMILES, graphs, 2D images, 3D coordinates, that characterize the structural information and serve as the inputs of prediction and generation. Knowledge-driven CAMD \cite{Odele1993,Churi1996,Camarda1999,Sinha1999,Sahinidis2003,Zhang2015,Liu2019,Alshehri2020,Hatamleh2022} commonly represents molecules by basic chemical units like atoms and bonds, which is more like a combination of fingerprints and graphs. Empirical score functions are usually designed to predict properties, and then optimized to get desired molecules. Data-driven CAMD \cite{Xu2017,Wang2019,Yang2019,Schweidtmann2020,Withnall2020,Alshehri2021,Sharma2021,Mann2023,Tiew2023,Rittig2022} usually has a generative and a predictive model, where the generative model maps between molecular representations and a latent space, and the predictive model evaluates the latent space and yields predictive properties. The latent space is then used as the domain of optimization. For more details about various CAMD methods, we refer to these reviews \cite{Gaudelet2021,Xiong2021,Elton2019,Xia2019,Alshehri2020,Faez2021,Gao2022,Tang2024}.

There is no clear boundary between knowledge-driven and data-driven CAMD. On the contrary, hybrid methods are proposed to benefit from both sides: (i) accumulated knowledge involved in molecular representations and latent space construction helps ML models learn more meaningful information, (ii) generative and predictive models overcome the computational limitations of traditional CAMD methods and efficiently exploit the uncharted chemical space. Here we highlight recent works \cite{Zhang2023,Zhang2024,McDonald2024} linking well-established mixed-integer programming (MIP) framework in knowledge-driven CAMD with optimization over trained ML models such as Gaussian processes \cite{Schweidtmann2021, Xie2024}, trees \cite{Misic2020,Mistry2021,Thebelt2021,Thebelt2022a,Thebelt2022b,Ammari2023}, neural networks (NNs) \cite{Tsay2021,Ceccon2022,Schweidtmann2022,Wang2023}, and GNNs \cite{Zhang2024,Hojny2024}. By encoding GNNs into optimization problems, these works directly optimize over the design space instead of latent space, and yield molecules with optimal predictive properties. The idea of incorporating property prediction into optimization of molecules is not new, which is already considered in CAMD packages such as AMODEO \cite{Samudra2013}, ProCAPD \cite{Kalakul2018} and OptCAMD \cite{Liu2019}. For example, the OptCAMD framework \cite{Liu2019} adds artificially-designed property constraints into their MIP formulation. But formulating predictive models as property objective/constraints shows a possibility of involving ML techniques in MIP-based CAMD \cite{Heng2022,Ooi2022,Chew2024}.

There are two major limitations to mixing MIP-based and data-driven CAMD. The first one is the weak scalability caused by the nonlinearity and nonconvexity of property constraints/predictive models. Although there are a few attempts to relieve this limitation, e.g., applying symmetry-breaking constraints to reduce design space \cite{Zhang2023}, extending atom-based design to fragment-based \cite{Liu2019,Zhang2024}, etc., the size of molecules yielded from these methods is still limited. Meanwhile, fragment-based methods require knowing all possible fragments as \emph{a prior} \cite{Jin2018,Jin2020,Podda2020,Green2021,Powers2022}, which also limits the diversity of molecules. The second one is the molecular feasibility, which is actually a general issue of CAMD. Here we borrow the criterion of reward function design in CAMD proposed by \citeauthor{Elton2019}\cite{Elton2019}: (i) diversity, (ii) novelty, (iii) stability, (iv) synthesizability, (v) non-trivility, and (vi) good properties. Beyond associating with a high reward, molecular feasibility could also either (i) be listed as constraints in design space, or (ii) be considered in an extra verification step after CAMD step. Incorporating feasibility constraints could help reduce the search space and avoid unreasonable molecules in advance, but complicated constraints, e.g., non-linear, sophisticated functions/models, could result in computationally expensive, even intractable optimization problems. Alternatively, considering molecular feasibility in verification step filters infeasible candidates, but there is no guarantee that there is a feasible molecule after verification.

This work is motivated by the potential and limitations of the hybrid methods based on MIP and ML: we take a step back and think how to exploit the applicability and compatibility of MIP framework. Our proposed framework, Limeade v1.0\footnote{\href{https://github.com/cog-imperial/Limeade}{https://github.com/cog-imperial/Limeade}, MIT License.}, is a lightweight, end-to-end generation tool, taking minimal information from users and generating feasible molecules quickly. By feasibility, we mean the mathematical feasibility of our optimization formulation, which is not saying that Limeade could generate practically desirable molecules. The principle of Limeade is using the feasible domain to approximate the desired molecular space as closely as possible by adding constraints: more constraints result in a smaller feasible domain and a larger probability of generating desired molecules. Although such principle is classic in MIP-based CAMD literature, the major contribution of Limeade is to bridge the gap between chemical engineering and mathematical programming: the interpretation and formulation of the molecular space and practical requirements are automatically processed into a MIP, then the feasible solutions by solving the MIP are decoded into molecules. Figure \ref{fig:Limeade} illustrates how Limeade works. Since all optimization-relevant procedures are hidden inside Limeade, users may apply this MIP-based molecular generation method without optimization expertise.  

\textbf{Paper structure} Section \ref{sec:molecular_encoding} introduces the integer molecular encoding framework, which is the foundation of Limeade. Section \ref{sec:library_design} shows all functionalities of Limeade with illustrative examples. Section \ref{sec:case_study} presents a case study where we use Limeade to iteratively generate feasible molecules with respect to Morgan fingerprints. Section \ref{sec:conclusion} summarizes and concludes.

\section{Integer molecular encoding}\label{sec:molecular_encoding}
Limeade is built on the well-established MIP encoding of molecules \cite{Odele1993,Churi1996,Camarda1999,Sinha1999,Sahinidis2003,Zhang2015,Liu2019,Alshehri2020,Hatamleh2022,Zhang2023,Zhang2024,McDonald2024}, and particularly inherits molecular representation and structural constraints from \citeauthor{Zhang2024}\cite{Zhang2024}. Beyond the basic feasibility considerations, we propose computationally tractable constraints to support application-specific preference on the substructure selection, i.e., including or excluding given substructures in generated molecules.

\subsection{Molecular representation}\label{subsec:molecular_representation}
We describe a molecule by its molecular graph structure and basic chemical units, i.e., atoms and bonds. For a molecule with $N$ heavy atoms (hydrogen atoms are considered implicitly), each atom is chosen from a given atom type set $Atom$. Let $N^t=|Atom|$ be the number of atom types, and $\mathcal C$ be the maximal covalence among all possible atoms, then each atom has $F$ features consisting of atom type (with size $N^t$), number of neighbors (with size $\mathcal C$), number of associated hydrogen atoms (with size $\mathcal C+1$), and types of adjacent bonds (with size $2$, i.e., double and triple bond). Gathering all atom features gives the feature matrix $X\in \{0,1\}^{N\times F}$ on atom level. On bond level, we use three adjacency matrices $A, DB, TB\in \{0,1\}^{N\times N}$ to express the relationships (arbitrary/double/triple bond, respectively) between any pair of atoms. The diagonal elements of $A$ could be treated as variables to handle changing number of atoms \cite{Zhang2023,Zhang2024,McDonald2024}, but we disable such functionality to reduce the design space to molecules with fixed size. Table \ref{table:summary_of_notations} summarizes the notation for building constraints and gives an example of our featurization.

    \begin{table}[t] 
        \caption{Summary of notations used in MIP formulation for CAMD. Their values correspond to the example with four types of atom $\{C,N,O,S\}$ \cite{Zhang2024}.}
        \label{table:summary_of_notations}
        \centering
        \vspace{1mm}
        \begin{tabular}{lll}
            \toprule   
            Symbol & Description & Value \\
            \midrule
            $N$ & number of heavy atoms &  *\\
            $F$ & number of features & $15$ \\
            $N^t$ & number of atom types & $4$ \\
            $N^n$ & number of neighbors & $4$ \\
            $N^h$ & number of hydrogen & $5$ \\
            \midrule
            $I^t$ & indexes for $N^t$ & $\{0,1,2,3\}$ \\
            $I^n$ & indexes for $N^n$ & $\{4,5,6,7\}$ \\
            $I^h$ & indexes for $N^h$ & $\{8,9,10,11,12\}$ \\
            $I^{db}$ & index for double bond & $13$ \\
            $I^{tb}$ & index for triple bond & $14$ \\
            \midrule
            $Atom$ & atom types &  $\{C,N,O,S\}$ \\ 
            $Cov$ & covalences of atoms & $\{4,3,2,2\}$ \\
            \bottomrule  
        \end{tabular}
    \end{table}

\subsection{Structural constraints}\label{subsec:structural_constriants}
Structural constraints are basic requirements that molecular candidates should satisfy. The constraints introduced here have previously appeared in the literature \citep{Odele1993,Churi1996,Camarda1999,Sinha1999,Sahinidis2003,Zhang2015,Zhang2023,Zhang2024,McDonald2024}. Therefore, we only use several representative constraints to illuminate how integer molecular encoding works, and report all constraints in Appendix \ref{app:full_constraints}. For simplicity, $[N]$ represents the set $\{0,1,\dots,N-1\}$.

Constraints \eqref{C3} require at least one bond between atom $v$ and any atom with index smaller than $v$, which guarantees the connectivity of any subgraph induced by atom set $\{0,1,\dots,v\}~,\forall v\in [N]$:
    \begin{align*}
        \sum\limits_{u<v} A_{u,v}&\ge 1,&&\forall v\in [N]\backslash\{0\}. \tag{C3}
    \end{align*}
Constraints \eqref{C9} exclude more than one atom type:
    \begin{align*}
        \sum\limits_{f\in I^t}X_{v,f}&=1,&&\forall v\in [N]. \tag{C9}
    \end{align*}
Constraints \eqref{C13} link double bond information between features and adjacent matrices, i.e., there is a double bond between atom $u$ and atom $v$ ($D_{u,v}=1$) if and only if bond atoms have double bond feature active ($X_{u,I^{db}}=X_{v, I^{db}}=1$) and they are linked ($A_{u,v}=1$):
    \begin{align*}
        3\cdot DB_{u,v}&\le X_{u,I^{db}}+X_{v,I^{db}}+A_{u,v},&&\forall u,v\in [N],u<v. \tag{C13}
    \end{align*}
Constraints \eqref{C19} are the covalence equations:
    \begin{align*}
        \sum\limits_{i\in[N^t]} Cov_i\cdot X_{v,I^t_i}&=\sum\limits_{i\in [N^n]} (i+1)\cdot X_{v,I^n_i} +\sum\limits_{i\in [N^h]}i\cdot X_{i,I^h_i} \\
            &+\sum\limits_{u\in[N]}DB_{u,v}+\sum\limits_{u\in[N]}2\cdot TB_{u,v},&&\forall v\in [N]. \tag{C19}
    \end{align*}

Constraints \eqref{C1} -- \eqref{C19} are compulsory for a valid molecule. Beyond them, one can optionally use bound constraints \eqref{C20} -- \eqref{C24} to control the composition of molecules, e.g., constraints \eqref{C20} bound the number of each type of atom:
    \begin{align*}
        \sum\limits_{v\in [N]}X_{v,I^t_i}&\in[LB_{Atom_i},~UB_{Atom_i}],&&\forall i\in [N^t]. \tag{C20}
    \end{align*}
    
For a molecular with $N$ heavy atoms, there are $N!$ ways to index $N$ atoms, that is, $N!$ different graph representations for the same molecule. This symmetry issue is a major challenge of molecular graph representations \cite{Elton2019}. \citeauthor{Zhang2023}\cite{Zhang2023} proposed symmetry-breaking constraints to resolve this issue without reducing the search space or violating other structural constraints. For example, constraints \eqref{C25} order atoms using lexicographic order:
\begin{align*}
        \sum\limits_{u\neq v,v+1}2^{N-u-1}\cdot A_{u,v}&\ge \sum\limits_{u\neq v,v+1}2^{N-u-1}\cdot A_{u,v+1},&&\forall v\in [N-1]\backslash\{0\}.\tag{C25}
\end{align*}

\subsection{Substructure inclusion \& exclusion}\label{subsec:substructure_inclusion_exclusion}
In practical molecular design, including certain substructures might be expected to have desired properties based on expert knowledge or experimental observations. Meanwhile, some substructures might be excluded because of instability, non-synthesizability, etc. This section defines both inclusion and exclusion constraints in the MIP framework and leverages the computational complexity for efficient generation.

Suppose we are either including or excluding a substructure with $n$ atoms with feature matrix $X_0\in\{0,1\}^{n\times F}$ and adjacency matrices $A_0, DB_0, TB_0\in\{0,1\}^{n\times n}$. To check if this substructure exists in the molecular candidate, we need to consider each substructure with equal size. Let $V=[N]$ be the atom set and $V_n=\{v_0,v_1,\dots,v_{n-1}\}\subset V$ be a subset of $V$ with size $n$. Assume that the feature matrix consisting of atoms $v_0,v_1\dots,v_{n-1}$ is $X_n\in\{0,1\}^{n\times F}$, and the adjacency matrices among these $n$ atoms are $A_n, DB_n, TB_n\in\{0,1\}^{n\times n}$. Since there is no extra information about the relationships between this substructure and the rest of the molecule, we can not just check if $(X_0,A_0,DB_0,TB_0)$ equals to $(X_n,A_n,DB_n,TB_n)$. Additionally, the given substructure could be a pattern, e.g., using SMARTS \cite{SMARTS2007} strings to give multiple options to each atom and bond. Therefore, instead of directly comparing features between the given and enumerated substructure, we consider the number of matched components between them. Table \ref{table:component_types} lists the types of components.

    \begin{table}[t] 
        \caption{Types of components of a substructure. The value of a component is $1$ if it satisfies the description, and $0$ otherwise.}
        \label{table:component_types}
        \centering
        \vspace{1mm}
        \begin{tabular}{ll}
            \toprule   
            Description & Component  \\
            \midrule
            arbitrary bond between atom $u$ and $v$ & $A_{u,v}$ \\
            no bond between atom $u$ and $v$ & $1-A_{u,v}$ \\
            single bond between atom $u$ and $v$ & $A_{u,v}-DB_{u,v}-TB_{u,v}$ \\
            double bond between atom $u$ and $v$ & $DB_{u,v}$ \\
            triple bond between atom $u$ and $v$ & $TB_{u,v}$ \\
            \midrule
            atom $v$ is chosen from set $\mathcal T\subset Atom$ & $\sum_{i\in\mathcal T}X_{v,i}$ \\
            \midrule
            atom $v$ has $d$ neighbors & $X_{v, N^t+d-1}$\\
            \midrule
            atom $v$ has $h$ associated hydrogen atoms & $X_{v,N^t+N^n+h}$ \\
            \bottomrule  
        \end{tabular}
    \end{table}
    
\textit{Remark:} There is more information that could be used in the substructure matching. But we choose the components in Table \ref{table:component_types} because (i) they are defined using the features introduced in Section \ref{subsec:molecular_representation}, (ii) they have linear forms, and (iii) they could be extracted from a SMARTS string using RDkit \cite{RDkit}.

Depending on the information received from the given substructure, the number of considered components might change. Assume $M$ components are involved, and denote the sum of all components from the substructure induce by $V_n$ is $\mathcal S(V_n)$. Mathematically, excluding the given substructure means:
    \begin{align}\label{eq:exclusion}
        \mathcal S(V_n)<M,~\forall V_n\subset V,
    \end{align}
while including the given substructure is equivalent to:
    \begin{align}\label{eq:inclusion_original_form}
        \exists V_n\subset V,~ s.t., \mathcal S(V_n)=M.
    \end{align}
    
The number of constraints in \eqref{eq:exclusion} equals to the number of ordered subsets of $V$, which is the permutation number $P(N,n)=\frac{N!}{(N-n)!}$. When $n\le 3$, adding $P(N,n)$ constraints is still tractable. For example, removing accumulations of heteroatoms like \texttt{S-S-S}, or allenes like \texttt{C=C=C}. However, with larger $N$ and $n$, excluding substructure using \eqref{eq:exclusion} involves too many constraints, e.g., $P(30,6)\approx 4.3\times 10^8, P(100,4)\approx 9.4\times 10^7$. Given that in these cases the probability of getting these substructures is small, we omit these requirements in our generation step and check them later in the validation step (see Section \ref{subsec:validation_step}). Without considering other $N-n$ atoms, the possibility is choosing one from all possible molecules with $n$ atoms. Based on the numerical results reported in \citeauthor{Zhang2023}\cite{Zhang2023}, the number of molecules with $6$ atoms (choosing from $4$ types of atom) after adding structural, bound, and symmetry-breaking constraints is around $5\times 10^4$, meaning the possibility is around $2\times 10^{-5}$. Such negligible possibility does not merit more efforts on the generation process: if the MIP generation procedure generates a molecule with an unwanted substructure, we just remove it from the candidates.

Including a substructure requires us to handle the disjunctive constraints in \eqref{eq:inclusion_original_form}. We introduce auxiliary variable $\sigma(V_n)\in\{0,1\}$ to indicate if $\mathcal S(V_n)=M$ using big-M:
    \begin{align}
        \mathcal S(V_n)\ge M\cdot \sigma(V_n),
    \end{align}
which implies:
    \begin{align}
        \sigma(V_n)=
        \begin{cases}
            1,&\mathcal S(V_n)=M\\
            0,&\mathcal S(V_n)<M
        \end{cases}.
    \end{align}
Then we can equivalently represent disjunctive constraints \eqref{eq:inclusion_original_form} as:
    \begin{equation}\label{eq:inclusion_bigM_form}
        \begin{aligned}
            \mathcal S(V_n)&\ge M\cdot \sigma(V_n),~\forall V_n\subset V,\\
        \sum\limits_{V_n\subset V}\sigma(V_n)&\ge 1.
        \end{aligned}
    \end{equation}

With larger $N$ and $n$, \eqref{eq:inclusion_bigM_form} is more intractable than \eqref{eq:exclusion} since it also involves $P(N,n)$ binary variables. But we cannot make the same trade-off as substructure exclusion and disregard substructure inclusion when $P(N,n)$ is large: generating several unwanted molecules is much less critical than when most generated molecules are unwanted. Recall that substructure inclusion is an existence problem, i.e., we do not need to generate all possible molecules having this given substructure. For instance, we can just fix the substructure consisting of the first $n$ atoms as the given one. However, this idea is not satisfactory when there are multiple substructures needed to be included and the relationships between these substructures are unclear. Therefore, we choose a relaxed version: the given substructure will appear in the resulting molecule and consist of $n$ atoms with consecutive indexes. Denote $V_n^i=\{i,i+1,\dots,n+i-1\},~0\le i\le N-n$. Then we use the following constraints to include a given substructure:
    \begin{equation}\label{eq:inclusion}
        \begin{aligned}
            \mathcal S(V_n^i)&\ge M\cdot \sigma(V_n^i),~0\le i\le N-n\\
            \sum\limits_{i=0}^{N-n}\sigma(V_n^i)&\ge 1
        \end{aligned}
    \end{equation}
which only adds $O(N-n)$ constraints and binary variables.

Note that substructure inclusion using \eqref{eq:inclusion} will hurt the symmetry-breaking constraints \eqref{C25} since we assign consecutive indexes to atoms belonging to the given substructure. Theoretically, it is still possible to apply symmetry-breaking constraints to other atoms: atom $v$ is not in the given substructure if and only $\sum_{|i-v|<n}\sigma(V_n^i)=0$, which can be used to restrict the comparison in \eqref{C25} to atoms not in the substructure. However, the practical implementation is not straightforward and complicates the encoding. Additionally, we have a similar observation as the substructure exclusion: the possibility of generating two symmetric molecules is quite small, and getting several symmetric candidates will not hurt much.

\section{Library Design}\label{sec:library_design}
This section introduces the implementation of Limeade with some illustrative examples. Limeade v1.0 uses Gurobi v11.0.0 \citep{Gurobi2023} to build and solve MIPs for molecular generation. By setting \texttt{PoolSearchMode=2}, Gurobi finds multiple feasible solutions to fill the solution pool, whose size \texttt{PoolSolutions} is set as the number of molecules needed. To additionally release Limeade to the open-source community, we also provide a Pyomo \citep{Pyomo2021} version of Limeade, thereby allowing use of open-source solvers. In this case, Limeade v1.0 uses Pyomo v6.8.1 and \texttt{pyomo.contrib.alternative\_solutions} function to generate feasible solutions.

\subsection{Basic functionalities}\label{subsec:basic_functionalities}
The core class \texttt{MIPMol} of Limeade takes a set of atom types \texttt{atoms} and the number of atoms \texttt{N\_atoms} as inputs, which correspond to $Atom$ and $N$ in Section \ref{subsec:molecular_representation}. Then Limeade computes all hyperparameters listed in Table \ref{table:summary_of_notations}. After that, \texttt{MIPMol} builds a Gurobi model with variables $X,A,DB,TB$ and structural constraints \eqref{C1} -- \eqref{C19}. Finally, given the number of molecules needed \texttt{NumSolutions}, this Gurobi model is solved and all feasible solutions will be retrieved and decoded into SMILES strings as outputs. The solving process will be terminated when (i) all feasible solutions are found, or (ii) the solution pool is full, or (iii) time out (the default time limit is $600$ seconds for each batch). See below for a demo of Limeade that uses only three lines Python code to finish all aforementioned procedures.

\begin{python}
# a three-line demo of Limeade
from Limeade import MIPMol
Mol = MIPMol(atoms=["C", "N", "O", "S"], N_atoms=10) # build the model
mols = Mol.solve(NumSolutions=100) # solve the model
\end{python}

The most important part of Limeade happens between building and solving the model, by enhancing the model with practical constraints. The following functions control the composition of molecules as shown in constraints \eqref{C20} -- \eqref{C23}, allowing users to bound the number of each type of atom, double/triple bounds, and rings.
\begin{python}
Mol.bounds_atoms(lb_atom, ub_atom) # bound the number of each type of atom
Mol.bounds_double_bonds(lb_db, ub_db) # bound the number of double bonds
Mol.bounds_triple_bonds(lb_tb, ub_tb) # bound the number of triple bonds
Mol.bounds_rings(lb_ring, ub_ring) # bound the number of rings
\end{python}

\subsection{Supported SMARTS patterns}\label{subsec:supported_SMARTS}
Those basic functionalities introduced in Section \ref{subsec:basic_functionalities} are already in the literature, but we aggregate and release these functionalities in open-source software. The key contributions of Limeade are substructure inclusion and exclusion, which enable users make requests using SMARTS \citep{SMARTS2007} strings without encoding their requirements into mathematical constraints. Limeade uses the Section \ref{subsec:substructure_inclusion_exclusion} constraints to include or exclude a substructure with components listed in Table \ref{table:component_types}. Those components correspond to SMARTS patterns as follows: 

\textbf{Atom-level patterns:} We can extract information of each atom from attribute \texttt{AtomType}. Three patterns are supported: (i) a single type, e.g., \texttt{C}, \texttt{N}, etc., (ii) multiple types, e.g., \texttt{[C, N]}, and (iii) arbitrary types, i.e., \texttt{*}. Note that all types must belong to \texttt{atoms}. For the last pattern, this atom will be set as any type from \texttt{atoms}. 

\textbf{Bond-level patterns:} Four types are considered: (i) single bond, i.e., \texttt{-}, (ii) double bond, i.e., \texttt{=}, (iii) triple bond, i.e., \texttt{\#}, and (iv) any type of bond, i.e., \texttt{$\sim$}. Note that aromatic bond is not considered. Therefore, any type of bond means one of the first three options.

\textbf{Graph-level patterns:} We also include two patterns on graph-level: (i) the number of implicit hydrogen atoms of each atom, which is given as attribute \texttt{AtomHCount}, and (ii) the number of neighbors of each atom, which is the difference between \texttt{AtomTotalDegree} and \texttt{AtomHCount}, where attribute \texttt{AtomTotalDegree} is the degree of each atom.

To show how to use Limeade to include and exclude patterns, we design an example where we try to generate aspirin ($\rm C_9H_8O_4$):
\begin{center}
    \includegraphics[width=0.2\textwidth]{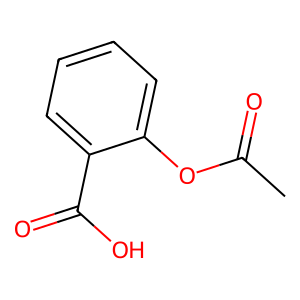}
\end{center}

\emph{Step 1:} Build a \texttt{MIPMol} class with $9$ carbon, $4$ oxygen, $5$ double bonds, and $1$ ring:
\begin{python}
Mol = MIPMol(atoms=["C", "O"], N_atoms=13) # 13 atoms chosen from [C, O]
Mol.bounds_atoms([9, 4], [9, 4]) # 9 carbon, 4 oxygen
Mol.bounds_double_bonds(5, 5) # 5 double bonds
Mol.bounds_triple_bonds(None, 0) # no triple bonds
Mol.bounds_rings(1, 1) # 1 rings
\end{python}

\begin{figure}[htp]
    \centering
    \begin{subfigure}{\textwidth}
        \centering
        \includegraphics[width=\linewidth]{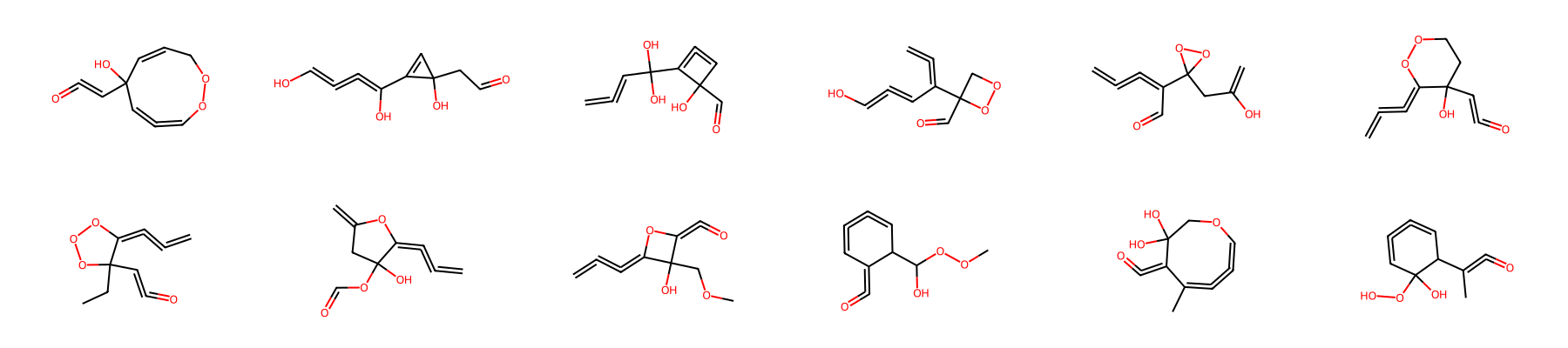}
        \caption{Step 1: fix the number of \texttt{C}, \texttt{O}, double/triple bonds, and rings.}
        \label{fig:aspirin_step_1}
    \end{subfigure}
    \vspace{3mm}
    
    \begin{subfigure}{\textwidth}
    \centering
        \includegraphics[width=\linewidth]{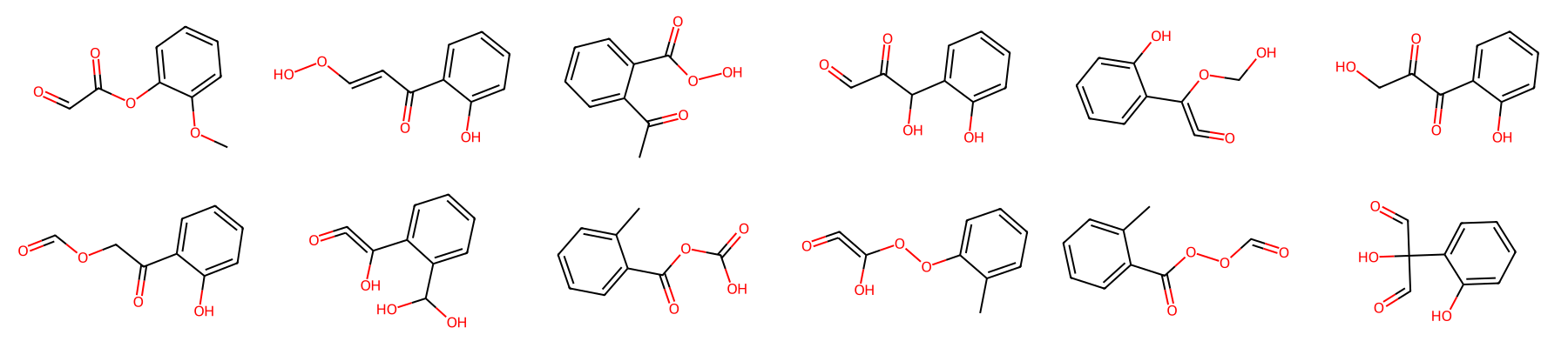}
        \caption{Step 2: include \texttt{C1=[CH][CH]=[CH][CH]=C1}, exclude \texttt{[CH]1=[CH][CH]=[CH][CH]=C1}.}
        \label{fig:aspirin_step_2}
    \end{subfigure}
    \vspace{3mm}
    
    \begin{subfigure}{\textwidth}
    \centering
        \includegraphics[width=\linewidth]{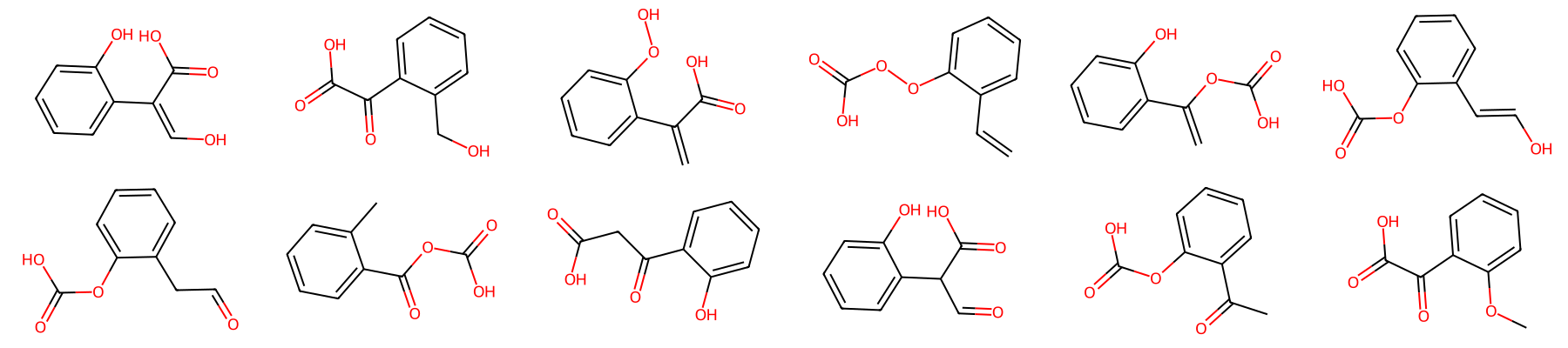}
        \caption{Step 3: include \texttt{O=C-[OH1]}.}
        \label{fig:aspirin_step_3}
    \end{subfigure}
    \vspace{3mm}
    
    \begin{subfigure}{\textwidth}
    \centering
        \includegraphics[width=\linewidth]{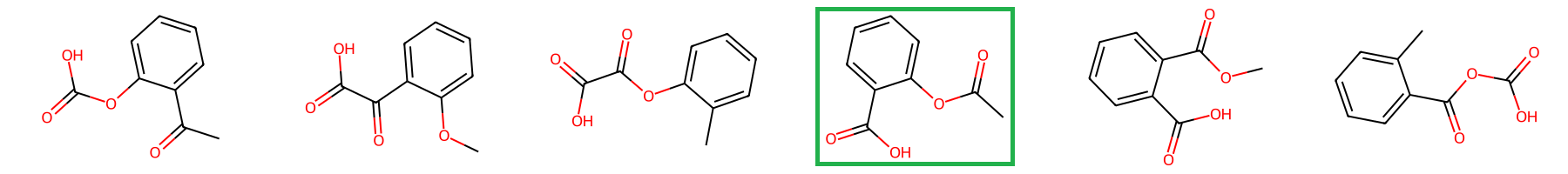}
        \caption{Step 4: include \texttt{[OH0X2]} and \texttt{[CH3]}.}
        \label{fig:aspirin_step_4}
    \end{subfigure}
    \caption{An example showing how to retrieve aspirin after including and excluding several patterns using Limeade. For each step, we generate $1000$ molecules and plot the first $10$ molecules. Note that there are only $6$ feasible molecules in the last step.}
    \label{fig:aspirin}
\end{figure}

Figure \ref{fig:aspirin_step_1} plots $10$ molecules, each of which has a random ring, while what we need is a benzene ring with two attachment points.

\emph{Step 2:} include \texttt{C1=[CH][CH]=[CH][CH]=C1}, exclude \texttt{[CH]1=[CH][CH]=[CH][CH]=C1}:
\begin{python}
# include benzene ring with two attachment points
Mol.include_substructures(["C1=[CH][CH]=[CH][CH]=C1"]) 
# exclude benzene ring with only one attachment point
Mol.exclude_substructures(["[CH]1=[CH][CH]=[CH][CH]=C1"])
\end{python}

Figure \ref{fig:aspirin_step_2} shows that only appears at a few molecules have carboxyl groups (\texttt{O=C-[OH1]}).

\emph{Step 3:} include \texttt{O=C-[OH1]}:
\begin{python}
Mol.include_substructures(["O=C-[OH1]"]) # include O=C-[OH1]
\end{python}

As shown in Figure \ref{fig:aspirin_step_3}, some molecules are already similar to aspirin. We identify two patterns in aspirin: (i) there is one oxygen atom with two neighbors, and (ii) there is one methyl group, i.e., carbon atom with three hydrogen atoms.

\emph{Step 4:} include \texttt{[OH0X2]} and \texttt{[CH3]}:
\begin{python}
Mol.include_substructures(["[OH0X2]", "[CH3]"]) # include [OH0X2], [CH3]
\end{python}

Eventually, we find aspirin within the remaining $6$ solutions. It is noteworthy that those constraints added above do not have practical meanings and the excluded substructures are probably needed in some real-world applications. This example is mainly used to show various SMARTS patterns that Limeade could interpret and formulate.

\subsection{Validation step}\label{subsec:validation_step}
Limeade generates mathematically feasible molecules. As long as a requirement is encoded into constraints, it will be satisfied by any molecule that Limeade generates. However, it is numerically difficult or intractable to encode all requirements. For example, numerous constraints need to be added to exclude a large substructure, as mentioned in Section \ref{subsec:substructure_inclusion_exclusion}. 

Moreover, as discussed in Section \ref{subsec:substructure_inclusion_exclusion}, substructure inclusion complicates the implementation of symmetry-breaking constraints \eqref{C24} -- \eqref{C25}. Since breaking symmetry for molecular generation is less critical than for optimal molecular design, Limeade does not apply symmetry-breaking techniques and symmetric molecules may appear in the solution pool. 

Because the solution pool may include symmetric molecules and molecules with substructures that the user wanted to exclude, Limeade adds an extra validation step after molecular generation to (i) remove symmetric molecules, and (ii) check if any molecule has unwanted substructures. All unwanted substructures are put in attribute \texttt{check\_later} of \texttt{MIPMol} class. For SMARTS strings that are not supported by Limeade, the users can directly add them into \texttt{check\_later}. For a supported SMARTS string, if too many constraints are needed to encode it, Limeade will also put it into \texttt{check\_later} instead of adding such many constraints.

With this validation step, the final number of molecules is usually less than \texttt{NumSolutions}, which is the number of molecules needed. Therefore, we recommend users to set a larger \texttt{NumSolutions} so that they can get enough unique and feasible molecules.

\subsection{Infeasibility detection}\label{subsec:infeasibility_detection}
Sometimes the users may not realize that some requirements contradict each other, which may result in infeasible models. To help users identify those contradictory requirements, Limeade uses Gurobi to compute an irreducible inconsistent subsystem (IIS) and identify problematic constraints. Since computing IIS is computationally expensive, Limeade will only do it when there are no more than $10^5$ constraints. For the current Pyomo version of Limeade, Limeade will only print an error message \texttt{Infeasible model}. See the following examples about this functionality.

\emph{Example 1:} Contradiction between bounds and substructure inclusion:
\begin{python}
Mol.bounds_rings(None, 0) # no rings
Mol.include_substructures(["C1=CC=CC=C1"]) # include a benzene ring
\end{python}
Error message from Limeade:
\begin{python}
Infeasible model. Please check the following constraints:
    -- upper bound of rings
    -- include C1=CC=CC=C1
\end{python}

\emph{Example 2:} Contradiction between substructure inclusion and exclusion:
\begin{python}
Mol.include_substructures(["S~N"]) # include S~N
Mol.exclude_substructures(["S-N", "S=N"]) # exclude S-N, S=N
\end{python}
Error message from Limeade:
\begin{python}
Infeasible model. Please check the following constraints:
    -- include S~N
    -- exclude S-N
    -- exclude S=N
\end{python}

\subsection{Batch-based generation}
The initial version of Limeade only solves the model once to generate all molecules needed by setting Gurobi parameter \texttt{PoolSolutions=NumSolutions}. However, we observe two undesired phenomena: (i) the generation time often grows much slower than linearly as the number of solutions increases, and (ii) there are too many symmetric solutions. To hasten the generation time and increase the diversity of solutions, Limeade v1.0 supports a batch-based generation. The users can set batch size when solving the model (the default is $100$), and Limeade will solve the model several times with different random seeds. For each batch, the default time limit is $600$ seconds, which the users could also change.

\section{Case study}\label{sec:case_study}
This section provides a full case study that uses Limeade to generate feasible molecules iteratively. In this case study, a molecule is labeled \emph{infeasible} if it contains a Morgan fingerprint with radius $1$ that occurs less than $5$ times in ChEMBL database \citep{ChEMBL2024}.

\emph{Iteration 1:} We first build a \texttt{MIPMol} class with $N=20$ atoms chosen from \texttt{[C, N, O, S]} and then set proper bounds to avoid molecules with extreme compositions:
\begin{python}
# set the number of atoms and types of atoms
N = 20
Mol = MIPMol(atoms=["C", "N", "O", "S"], N_atoms=N)
# set the bounds for the number of each type of atom
lb = [N // 2, None, None, None]
ub = [None, N // 4, N // 4, N // 4]
Mol.bounds_atoms(lb, ub)
# set the bounds for number of double/triple bonds, and rings
Mol.bounds_double_bonds(None, N // 2)
Mol.bounds_triple_bonds(None, N // 2)
Mol.bounds_rings(None, 0)
\end{python}

Then we use Limeade to generate $1000$ molecules, check their feasibility, and count the number of uncommon patterns, i.e., Morgan fingerprints with radius $1$. Without excluding any substructures, Limeade generates $957$ unique molecules, only $2$ of which are feasible. The top-10 uncommon patterns are listed as follows in form (\texttt{SMARTS}, \texttt{Count}):
\begin{python}
("CC(C)(C)N",717) ("CC(C)(C)C",309) ("CC(C)(N)O",282) ("CC(C)(N)N",213)
("CC(C)(C)O",194) ("CC(C)(C)S",185) ("CC(C)(O)S",177) ("CC(C)(N)S",155)
("CC(C)(O)O",99) ("C=C(C)C",85)
\end{python}

\emph{Iteration 2:} From those uncommon patterns obtained in \emph{Iteration 1}, we notice that most molecules are infeasible since they contain a carbon without any implicit hydrogen atom. Let us exclude this substructure and check if we can get more feasible molecules:
\begin{python}
Mol.exclude_substructures(["[CH0]"]) # exclude [CH0]
\end{python}
This time Limeade gives $970$ unique molecules, $62$ of which are feasible. The top-10 uncommon patterns are:
\begin{python}
("CN(C)N",427) ("CC(N)N",361) ("CC(N)O",195) ("CC(N)S",182) ("CN(N)N",178)
("CN(N)O",177) ("CN(C)S",106) ("CSN",93) ("CN=S",84) ("CN(C)O",69)
\end{python}

\emph{Iteration 3:} We indeed get more feasible molecules at \emph{Iteration 2} than \emph{Iteration 1}, but most molecules are still infeasible due to heteroatom-heteroatom or a carbon linked with two heteroatoms. In this iteration, we exclude both substructures:
\begin{python}
# exclude [N,O,S]~[N,O,S] and [N,O,S]~C~[N,O,S]
Mol.exclude_substructures(["[N,O,S]~[N,O,S]", "[N,O,S]~C~[N,O,S]"])
\end{python}
After this iteration, $787$ out of $855$ unique molecules generated from Limeade are feasible!

To test if we could achieve similar performance in larger scale, we use the same constraints to generate $10^5$ molecules. Limeade finds $63,973$ unique molecules and $55,498$ of them (around $86\%$) are feasible.

\section{Conclusion}\label{sec:conclusion}
We propose Limeade, a MIP framework for molecular generation, that automatically encodes molecular space and practical requirements. Limeade supports substructure inclusion and exclusion feeding in SMARTS patterns varying from atom-level to graph-level. Incompatible patterns could also be added into Limeade and will be checked in a validation step after the generation process. To enhance the user experience, Limeade will automatically remove symmetric solutions, and provide informative message when the model is infeasible. Benefiting from batch-based generation, both the generation speed and the diversity of solutions are increased. Multiple examples, as well as a case study, show the efficiency and potential of Limeade as a plug-and-play generation tool.

\begin{acknowledgement}

This work was supported by the Engineering and Physical Sciences Research Council [grant number EP/W003317/1], BASF SE, Ludwigshafen am Rhein to SZ, and a BASF/RAEng Research Chair in Data-Driven Optimisation to RM. We would like to thank Benjamin Merget, Jochen Sieg, Long Wang, Niklas Bjoern Wulkow and Philipp Eiden for their great comments and suggestions, which are very helpful to improve Limeade. We also appreciate the help from Yilin Xie and Yidi Shang for designing the Limeade logo.

\end{acknowledgement}

\newpage
\appendix

\begin{suppinfo}


\section{Full structural constraints}\label{app:full_constraints}
We report all constraints used in Limeade for structural feasibility.
\subsection{Feasible graph structure}
Constraints \eqref{C1} and \eqref{C2} force a symmetric adjacency matrix with fixed diagonal elements. Constraints \eqref{C3} require at least one bond between atom $v$ and any atom with index smaller than $v$, which guarantees the connectivity of any subgraph induced by atom set $\{0,1,\dots,v\}~,\forall v\in [N]$.
    \begin{align*}
        A_{v,v}&=1,&&\forall v\in [N] \label{C1}\tag{C1} \\
        A_{u,v}&=A_{v,u},&&\forall u,v\in [N],~u<v \label{C2}\tag{C2} \\
        \sum\limits_{u<v} A_{u,v}&\ge 1,&&\forall v\in [N]\backslash\{0\} \label{C3}\tag{C3}
    \end{align*}

\subsection{Feasible bond features}
Constraints \eqref{C4} -- \eqref{C7} define symmetric double/triple bond matrix with zero diagonal elements. \eqref{C8} restrict bond type when a bond exists.
    \begin{align*}
        DB_{v,v}&=0,&&\forall v\in [N] \label{C4}\tag{C4} \\
        DB_{u,v}&=DB_{v,u},&&\forall u,v\in [N],~u<v \label{C5}\tag{C5} \\
        TB_{v,v}&=0,&&\forall v\in [N] \label{C6}\tag{C6} \\
        TB_{u,v}&=TB_{v,u},&&\forall u,v\in [N],~u<v \label{C7}\tag{C7} \\
        DB_{u,v}+TB_{u,v}&\le A_{u,v},&&\forall u,v\in [N],~u<v \label{C8}\tag{C8} 
    \end{align*}

\subsection{Feasible atom features}
Constraints \eqref{C9} -- \eqref{C11} exclude more than one atom type, fix the number of neighbors, and set the number of associated hydrogen atoms.
    \begin{align*}
        \sum\limits_{f\in I^t}X_{v,f}&=1,&&\forall v\in [N] \label{C9}\tag{C9} \\
        \sum\limits_{f\in I^n}X_{v,f}&=1,&&\forall v\in [N] \label{C10}\tag{C10} \\
        \sum\limits_{f\in I^h}X_{v,f}&=1,&&\forall v\in [N] \label{C11}\tag{C11}
    \end{align*}

\subsection{Compatibility between atoms and bonds}
Constraints \eqref{C12} match the number of neighbors calculated from the adjacency matrix and atom features. \eqref{C13} -- \eqref{C18} consider the compatibility between atom features and bond features. \eqref{C19} correspond to the covalence equations.
{\allowdisplaybreaks
    \begin{align*}
        \sum\limits_{u\neq v}A_{u,v}&=\sum\limits_{i\in [N^n]}(i+1)\cdot X_{v,I^n_i},&&\forall v\in [N] \label{C12}\tag{C12} \\
        3\cdot DB_{u,v}&\le X_{u,I^{db}}+X_{v,I^{db}}+A_{u,v},&&\forall u,v\in [N],u<v \label{C13}\tag{C13} \\
        3\cdot TB_{u,v}&\le X_{u,I^{tb}}+X_{v,I^{tb}}+A_{u,v},&&\forall u,v\in [N],u<v \label{C14}\tag{C14} \\
        \sum\limits_{u\in [N]} DB_{u,v}&\le \sum\limits_{i\in[N^t]} \left\lfloor\frac{Cov_i}{2}\right\rfloor \cdot X_{v,I^t_i},&&\forall v\in [N] \label{C15}\tag{C15} \\
        \sum\limits_{u\in [N]} TB_{u,v}&\le \sum\limits_{i\in[N^t]} \left\lfloor\frac{Cov_i}{3}\right\rfloor \cdot X_{v,I^t_i},&&\forall v\in [N] \label{C16}\tag{C16} \\
        X_{v,I^{db}}&\le \sum\limits_{u\in [N]}DB_{u,v},&&\forall v\in [N] \label{C17}\tag{C17} \\
        X_{v,I^{tb}}&\le \sum\limits_{u\in [N]}TB_{u,v},&&\forall v\in [N] \label{C18}\tag{C18} \\
        \sum\limits_{i\in[N^t]} Cov_i\cdot X_{v,I^t_i}&=\sum\limits_{i\in [N^n]} (i+1)\cdot X_{v,I^n_i} +\sum\limits_{i\in [N^h]}i\cdot X_{i,I^h_i} \\
            &+\sum\limits_{u\in[N]}DB_{u,v}+\sum\limits_{u\in[N]}2\cdot TB_{u,v},&&\forall v\in [N] \label{C19}\tag{C19}
    \end{align*}
}
    
\subsection{Avoiding spurious extrapolation}
Constraints \eqref{C20} -- \eqref{C23} bound the number of each type of atom, double/triple bonds, and rings using observations from the dataset. By setting proper bounds, we can control the composition of the molecule, and avoid extreme cases such as all atoms being set to oxygen, or a molecule with too many rings or double/triple bounds.
{\allowdisplaybreaks
    \begin{align*}
        \sum\limits_{v\in [N]}X_{v,I^t_i}&\in[LB_{Atom_i},~UB_{Atom_i}],&&\forall i\in [N^t] \label{C20}\tag{C20} \\
        \sum\limits_{v\in [N]}\sum\limits_{u<v}DB_{u,v}&\in[LB_{db},~UB_{db}] && \label{C21}\tag{C21} \\
        \sum\limits_{v\in [N]}\sum\limits_{u<v}TB_{u,v}&\in [LB_{tb},~UB_{tb}] && \label{C22}\tag{C22} \\
        \sum\limits_{v\in [N]}\sum\limits_{u<v}A_{u,v}-(N-1)&\in [LB_{ring},~UB_{ring}] && \label{C23}\tag{C23}
    \end{align*}
    }

\subsection{Symmetry-breaking constraints}

Constraints \eqref{C24} force atom $0$ to have the most special features compared to other atoms, where a hierarchical linear function with coefficients $2^{F-f-1},f\in [F]$ is designed to map atom features to an integer in $[0,2^F-1]$. \eqref{C25} order all atoms except for atom $0$ by requiring that node $v$ should have a neighbor set with smaller lexicographic order compared to node $v+1$.
\begin{align*}
        \sum\limits_{f\in [F]}2^{F-f-1}\cdot X_{0,f}&\le \sum\limits_{f\in [F]} 2^{F-f-1}\cdot X_{v,f},&&\forall v\in [N]\backslash\{0\}\label{C24}\tag{C24}\\
        \sum\limits_{u\neq v,v+1}2^{N-u-1}\cdot A_{u,v}&\ge \sum\limits_{u\neq v,v+1}2^{N-u-1}\cdot A_{u,v+1},&&\forall v\in [N-1]\backslash\{0\}\label{C25}\tag{C25}
\end{align*}

\end{suppinfo}

\bibliography{ref}

\end{document}